# Efficient field-free perpendicular magnetization switching by a magnetic spin Hall effect

Shuai Hu[1,†], Ding-Fu Shao[2,†], Huanglin Yang[1,†], Meng Tang[1], Yumeng Yang[3,*], Weijia Fan[1], Shiming Zhou[1], Evgeny Y. Tsymbal[2,*] and Xuepeng Qiu[1,*]

[1]Shanghai Key Laboratory of Special Artificial Microstructure Materials and Technology and School of Physics Science and Engineering, Tongji University, Shanghai 200092, China

[2]Department of Physics and Astronomy and Nebraska Center for Materials and Nanoscience, University of Nebraska, Lincoln, Nebraska 68588-0299, USA

[3]School of Information Science and Technology, ShanghaiTech University, Shanghai 201210, China

[†]These authors contributed equally to this work.
[*]Email: yangym1@shanghaitech.edu.cn, tsymbal@unl.edu & xpqiu@tongji.edu.cn

**Current induced spin-orbit torques driven by the conventional spin Hall effect are widely used to manipulate the magnetization. This approach, however, is nondeterministic and inefficient for the switching of magnets with perpendicular magnetic anisotropy that are demanded by the high-density magnetic storage and memory devices. Here, we demonstrate that this limitation can be overcome by exploiting a magnetic spin Hall effect in noncollinear antiferromagnets, such as $Mn_3Sn$. The magnetic group symmetry of $Mn_3Sn$ allows generation of the out-of-plane spin current carrying spin polarization induced by an in-plane charge current. This spin current drives an out-of-plane anti-damping torque providing deterministic switching of perpendicular magnetization of an adjacent Ni/Co multilayer. Compared to the conventional spin-orbit torque devices, the observed switching does not need any external magnetic field and requires much lower current density. Our results demonstrate great prospects of exploiting the magnetic spin Hall effect in noncollinear antiferromagnets for low-power spintronics.**

A spin-orbit torque (SOT) provides an effective approach to manipulate the magnetization in spintronic devices[1-5]. In a typical SOT device, an in-plane charge current $\boldsymbol{J}$ generates a spin polarization $\boldsymbol{p}$ via spin-orbit coupling and exerts a torque on the magnetization $\boldsymbol{m}$ of a neighboring ferromagnetic layer. The most efficient SOT driven magnetization switching would require $\boldsymbol{p}$ to be parallel to the easy axis of $\boldsymbol{m}$, so that its associated anti-damping torque $\sim \boldsymbol{m} \times (\boldsymbol{m} \times \boldsymbol{p})$ can directly change the effective damping of a ferromagnet (FM)[6]. In the conventional spin Hall effect (SHE)[3,4] or the Rashba-Edelstein effect (REE)[1,2], the induced spin polarization $\boldsymbol{p}$, either in a heavy metal (HM) or at a nonmagnetic metal (NM)/FM interface, is always aligned the in-plane direction determined by $\boldsymbol{J} \times \boldsymbol{z}$ (where $\boldsymbol{z}$ is a normal to the film plane)[7,8]. The associated in-plane anti-damping torque is thus favorable to switch a magnet with in-plane magnetization, while its use for the switching of a magnet with perpendicular magnetization is nondeterministic[6]. Although this deficiency could be alleviated by applying an external assisting magnetic field and a high current density (Fig. 1a), these additional requirements would eventually hinder the application of the SOT technique in high-density and low-power spintronic devices.

A more efficient approach is to exploit an out-of-plane anti-damping torque $\sim \boldsymbol{m} \times (\boldsymbol{m} \times \boldsymbol{z})$ that directly counteracts the magnetization damping in a perpendicular magnet (Fig. 1b)[9-12]. Recently, extensive studies have been aimed at creating $z$-polarized spins that are capable to produce this kind of SOTs[13]. In particular, it has been shown that using a spin source with low crystal symmetry[14,15] and/or an additional magnetic order[15-18] can give rise to the $z$-polarized spins that are not allowed to appear in the conventional HM spin sources. Alternatively, a proper interface engineering has been reported as a possible mechanism to polarize spins along the $z$-direction through combined actions of spin-orbit filtering, spin precession, and scattering[19-24]. In terms of practical device application, it would be more straightforward and reliable to optimize the bulk property of the spin-source layer rather than to modify the interface between different layers. So far, however, switching of a perpendicular magnet has only been experimentally realized in FM/NM/FM trilayers, where the out-of-plane anti-damping torque was induced by the FM/NM interface[24].

Here, we demonstrate that an out-of-plane anti-damping torque can be directly generated in a SOT device based on the noncollinear antiferromagnet $Mn_3Sn$ due to the magnetic spin Hall effect[25,26], thus providing an efficient mechanism for perpendicular magnetization switching. We show that this switching is deterministic even in the absence of an applied magnetic field and requires a much lower critical current density than that based on the conventional SHE.

Bulk $Mn_3Sn$ is a hexagonal compound with the $Ni_3Sn$-type crystal structure[27]. As depicted in Figure 1c, Mn atoms form Kagome-type lattice planes stacked along the *c* axis, whereas Sn atoms are located at the center of the Mn-hexagons. The frustration from the triangular geometry of the Kagome lattice results in the chiral alignment of the Mn moments within each plane, leading to a noncollinear antiferromagnetic order of $Mn_3Sn$ with the Néel temperature ($T_N$) of ~420 K[28-30]. The noncollinear antiferromagnetism in $Mn_3Sn$ results in many interesting properties, such as the large room temperature anomalous Hall effect[27], the Weyl semimetal phase[31], and the magnetic spin Hall effect (MSHE)[32] relevant to our studies. Different from the conventional SHE being even under time reversal symmetry, the MSHE is odd with respect to this symmetry and hence reversable by flipping the magnetic moments. Being a consequence of symmetry breaking caused by the noncollinear magnetic structure, the MSHE can be intuitively regarded as a "magnetic" version of the SHE[25,33,34]. The MSHE gives rise to the unconventional spin currents with distinct symmetries providing useful implications for SOT devices. For example, an out-of-plane spin current with a finite *z*-spin component can be induced by the in-plane charge current (along the *x* or *y* direction) in the $Mn_3Sn$ (0001) film (Fig. 1c). This spin current is related to the magnetic spin Hall conductivity (MSHC) $\sigma_{zx}^z$ or $\sigma_{zy}^z$ (in the form of $\sigma_{jk}^i$, where $i$, $j$ and $k$ are the spin polarization, spin-current and charge-current directions, respectively), whose spin polarization $\boldsymbol{p}$ can be reversed by flipping the magnetic moments in $Mn_3Sn$. There are two types of the inverse triangular alignment of the magnetic moments in $Mn_3Sn$, denoted as AFM1 and AFM2 in Figure 1c. The magnetic group symmetry of these alignments supports a small but nonvanishing in-plane net magnetizations along the *x* ($[01\bar{1}0]$) direction for AFM1 and along the *y* ($[2\bar{1}\bar{1}0]$) direction for AFM2[28,29]. The presence of this magnetization allows, in particular, the control of antiferromagnetic domains by a small in-plane magnetic field[28]. This is reflected by our first-principles calculation (Fig. 1d), showing that the $\sigma_{zx}^z$ and $\sigma_{zy}^z$ are finite and reversable for AFM1 and AFM2 of $Mn_3Sn$, respectively. These sizable MSHCs allow using $Mn_3Sn$ as a spin source material to generate the out-of-plane anti-damping SOT and switch the magnetization of an adjacent ferromagnet.

On this basis, we design a MSHE SOT device consisting of a crystalline Mn$_3$Sn (0001) film to generate the MSHE, a [Ni/Co]$_3$ multilayer with perpendicular magnetic anisotropy to control its magnetization, and a thin Cu spacer layer to magnetically decouple Mn$_3$Sn and [Ni/Co]$_3$. Figure 2a schematically shows a stack of Mn$_3$Sn(7)/Cu(1)/[Ni(0.4)/Co(0.2)]$_3$ (number in the parentheses indicate thickness in nanometers) is epitaxially prepared by dc magnetron sputtering on cubic MgO (111) substrate. Hall bar devices are subsequently fabricated by standard lithography and etching process for electrical measurements. Through combined characterization techniques of X-ray diffraction, high-resolution transmission electron microscopy, and magneto-transport measurements (Supplementary Information S2, S3), the Mn$_3$Sn film is confirmed to have well-defined (0001) orientation, and its quality is comparable to the previously fabricated bulk Mn$_3$Sn crystal[31] and sputtered epitaxial films[35-37]. Strong perpendicular anisotropy of the [Ni/Co]$_3$ multilayer is confirmed by anomalous Hall effect (AHE) measurements (Supplementary Information S4).

We first verify the existence of the *z*-polarized component in the spin current by measuring the hysteresis loop of the anomalous Hall resistance $R_{AHE}$ versus the out-of-plane magnetic field $H_z$ in the presence of a bias dc current *I* along the *x* ([01$\bar{1}$0]) direction. If there was a *z*-polarized spin component, the associated out-of-plane anti-damping torque would cause an abrupt shift of the $R_{AHE}$-$H_z$ hysteresis loop when the torque is sufficiently strong to overcome the intrinsic damping of the Ni/Co multilayer[24]. Indeed, we find no shift of the loop when the amplitude of *I* is at 4 mA (Fig. 2b), while a sizable positive or negative shift occurs when *I* = +16 mA or −16 mA, respectively (Fig. 2c). Figure 2d, shows that the threshold current to produce the $R_{AHE}$-$H_z$ hysteresis loop shift is around ~10 mA, above which the shift increases almost linearly with the increase of *I*. A similar current induced shift and threshold effect are also observed when *I* is applied along the *y* ([2$\bar{1}\bar{1}$0]) direction (Supplementary Information S5).

The emergence of the *z*-polarized spin Hall current in Mn$_3$Sn allows the realization of the deterministic field-free magnetization switching. Figure 3a shows that the measured AHE resistance as a function of a pulse current *I* applied along the [01$\bar{1}$0] direction exhibits a hysteretic behavior and a sign change in the absence of applied magnetic field. This behavior reflects magnetization reversal in the Ni/Co multilayer. We note that in our SOT device, the exchange coupling between the Mn$_3$Sn and Ni/Co layers is eliminated by the Cu spacer, and hence it does not contribute to the magnetization switching. Instead, the field-free switching is entirely caused by the out-of-plane anti-damping torque. Furthermore, we find that about 60% of the magnetic Ni/Co layer volume switches as estimated from the measured $R_{AHE}$-$I$ loop (Fig. 3a) with respect to the saturation AHE resistance. This partial switching behavior is due to the multi-domain structure of the AFM Mn$_3$Sn film. Due to small magnetic anisotropy and nearly identical energies of AFM1 and AFM2 magnetic orders, as-grown Mn$_3$Sn films exhibit multiple antiferromagnetic domains which have a tendency to align randomly in zero field. This gives rise to variation in magnitude and sign of MSHC $\sigma_{zx}^z$ among different domains, and eventually leads to a reduced effective $\sigma_{zx}^z$. Therefore, the switching ratio is expected to increase if we align the AFM domains. This can be done by applying a finite magnetic field due to a small in-plane magnetization in Mn$_3$Sn[28]. Figure 3b shows the measured $R_{AHE}$-$I$ loops for different magnetic fields $H_x$ along the [01$\bar{1}$0] direction, and Figure 3c displays the corresponding switching ratio as a function of $H_x$. As seen, by aligning more domains into one configuration, the switching ratio increases as the magnitude of the field increases, with a maximum value of ~80% at a field as small as ~30 Oe. Importantly, the switching polarity is reversed upon reversing the field to opposite direction. All these observations corroborate with the MSHE scenario showing that the MSHC $\sigma_{zx}^z$ is controlled by the reorientation of Mn$_3$Sn domains[35, 38]. It is notable that flipping the switching polarity, accompanied by a vanished switching ratio, does not appear at zero field but at a small negative field of about −8 Oe (Fig. 3c). It is likely that some preferred AFM domain orientation is induced by structural reconstructions at interfaces or near defects during the deposition process, which as if there were a small bias field (~−8 Oe) on Mn$_3$Sn. Similar behavior is found when the current and magnetic field are both along the $y$ ([2$\bar{1}\bar{1}$0]) direction (Supplementary Information S5).

Finally, we compare the SOT switching efficiency of the Mn$_3$Sn based MSHE device with a conventional $β$-Ta based SHE device. The two devices have the same geometry and thickness of the corresponding layers (Figs. 4a and 4b). Here, a thicker 2 nm Cu layer is used in both devices as this is the minimum thickness required to achieve good perpendicular anisotropy in a [Ni/Co]$_3$ multilayer for the $β$-Ta based device. As a widely explored spin source, $β$-Ta has a large charge-spin conversion efficiency to generate the spin current with a conventional in-plane spin polarization[4]. This source, however, requires an assistive magnetic field applied parallel to the charge current for deterministic switching. We find that only a small switching ratio of ~17% is realized in this device at the sacrifice of a large current density of $1.3 \times 10^7$ A cm$^{-2}$ and a moderate field of 300 Oe (Fig. 4a). On the contrary, a much larger switching ratio of ~60% is achieved in the Mn$_3$Sn based MSHE device even in the absence of external field and under application of a much lower current density of $4.6 \times 10^6$ A cm$^{-2}$ (Fig. 4b). This evidence unambiguously proves the superior efficiency of the MSHE induced out-of-plane anti-damping torque for the deterministic switching of perpendicular magnetization.

In conclusion, we have demonstrated the efficient current-induced field-free switching of a ferromagnet with perpendicular magnetization. The magnetization switching is driven by an out-of-plane anti-damping SOT generated by non-collinear antiferromagnet Mn$_3$Sn resulting from the MSHE. Due to the magnetic spin Hall current being collinear to its spin polarization, this observed mechanism of switching requires much lower critical current density compared to the conventional SHE based SOT devices and allows control through the reorientation of magnetic domains in Mn$_3$Sn. Our findings pinpoint the enormous potential of the MSHE as the spin-torque source to engineer novel energy-efficient spintronic devices.

## Methods

First-principles density functional theory (DFT) calculations were performed using a plane-wave pseudopotential method with the fully-relativistic ultrasoft pseudopotentials[39] implemented in Quantum-ESPRESSO[40]. The exchange and correlation effects were treated within the generalized gradient approximation (GGA)[41]. In the calculations, we used the plane-wave cut-off energy of 35 Ry and a 16 × 16 × 16 k-point mesh in the irreducible Brillouin zone. The experimental lattice parameters $a$ = 5.689 Å and $c$ = 4.522 Å measured in this work were used in the calculation. All the atomic co-ordinates were relaxed until the force on each atom was less than 0.001 eV/Å. In the calculations, we firstly set the initial magnetic configurations according to the experimentally observed noncollinear antiferromagnetic states, and then performed full relaxations of the magnetic structure, and the electronic structure without any constraint.

The tight-binding Hamiltonians were obtained from the maximally localized Wannier functions[42] within the Wannier90 code[43] to calculate the magnetic spin Hall conductivity (MSHC)[25]

$$\sigma_{ij}^k = -\frac{e\hbar}{\pi} \int \frac{d^3\vec{k}}{(2\pi)^3} \sum_{n,m} \frac{\Gamma^2 \text{Re}(\langle n\vec{k}|J_i^k|m\vec{k}\rangle\langle m\vec{k}|v_j|n\vec{k}\rangle)}{\left[(E_F - E_{n\vec{k}})^2 + \Gamma^2\right]\left[(E_F - E_{m\vec{k}})^2 + \Gamma^2\right]}, \quad (1)$$

where $J_i^k = \frac{1}{2}\{v_i, s_k\}$ is the spin-current operator, $v_i$ and $s_k$ are velocity and spin operators, respectively, and $i, j, k = x, y, z$. $\Gamma = 50$ meV[25] and a $200 \times 200 \times 200$ $k$-point mesh were used to evaluate the integral of Eq. (1).

The symmetry determined geometries of conductivity tensors were obtained using the linear response symmetry code[44].

**Sample preparation.** All samples were deposited on MgO (111) substrates using DC magnetron sputtering with a base pressure of $5\times10^{-8}$ Torr. $Mn_3Sn$ films were deposited with a stoichiometric target at the Ar pressure of $2\times10^{-3}$ Torr, the deposition was performed at 100 °C, followed by annealing at 400 °C for 1h. After the $Mn_3Sn$ film cooled down to the room temperature, Cu and Ni/Co multilayers were then deposited on the $Mn_3Sn$ film with a Cu, Ni, and Co target, respectively. Standard photolithography and Ar ion etching were used to fabricate the 8 μm wide and 35μm long Hall bar.

**Sample characterization.** Structural properties of the samples were characterized using a Bruker D8 Discover X-ray diffraction (XRD) system with Cu Kα radiation. The HR-TEM were performed with an electron microscope operated at 200 kV (FEI Titan Themis 200). For current induced magnetization switching measurements, current pulses with a constant duration of 800 μs but varying amplitude were applied to the Hall bar. In-between two adjacent writing pulses, the magnetization state of the Ni/Co multilayer was read by current pulses with the same duration as that of the writing pulse but at a much smaller amplitude of 1 mA.

**Data availability**

The data that support the findings of this study are available from the corresponding author upon reasonable request.

**Acknowledgements**

This work was supported by the National Key R&D Program of China Grand No. 2017YFA0303202 and 2017YFA0305300, the National Natural Science Foundation of China Grant Nos. 52022069, 11974260, 11674246, 51501131, 51671147, 11874283, 51801152, and 11774064, Natural Science Foundation of Shanghai Grant No. 19ZR1478700, and the Fundamental Research Funds for the Central Universities.

**Author contributions**

S.H. and X.Q. conceived and designed the experiment. S.H. and H.Y. fabricated the samples and performed the measurements. S.H., D.S., H.Y., M.T., Y.Y., W.F., S.Z. and X.Q. analyzed and discussed the experiment results. F.S., E.Y.T. analyzed the data and performed the numerical calculation. S.H., Y.Y. wrote the manuscript with contributions from all the authors. All authors discussed the results and commented on the manuscript.

**Competing interests**

The authors declare no competing interests.

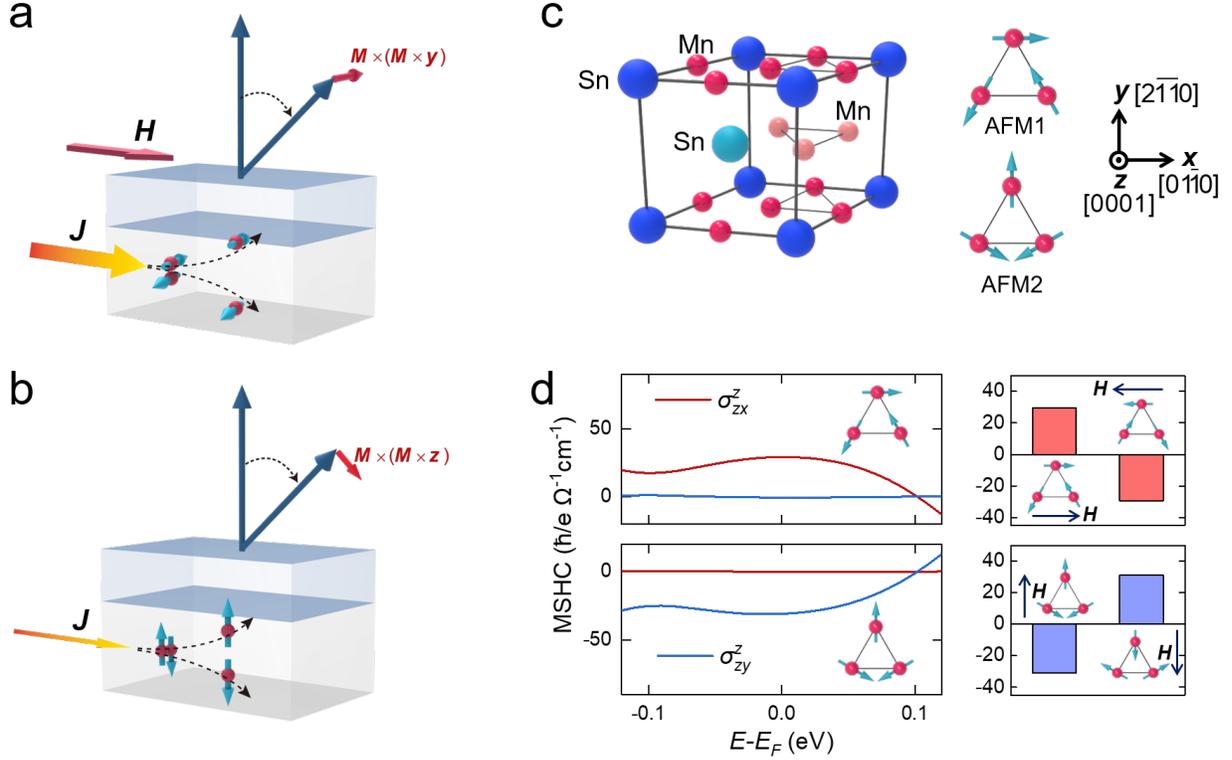

**Figure 1: Switching of perpendicular magnetization by damping-like SOTs. a** A schematic of a conventional bilayer SOT device. An in-plane charge current passes along *x* direction in the bottom spin source layer, generates an out-of-plane spin current with ***y*-polarized spin** through SHE. This spin current exerts an in-plane damping-like torque $\sim m \times m \times y$ on the perpendicular magnetization in the top ferromagnetic layer. In this case, a sizable external magnetic field is required for a deterministic switching, and the charge current required is large. **b** A schematic of a bilayer SOT device supporting the out-of-plane anti-damping torque, where the in-plane charge current generates an out-of-plane spin current with ***z*-**polarized spin. This spin current exerts an out-of-plane anti-damping torque $\sim m \times m \times z$ on the perpendicular magnetization in the top layer to realize a field-free switching, which does not require a large charge current. **c** The structure of the noncollinear antiferromagnetic $Mn_3Sn$. The left panel is the side view of the unit cell. The right panel shows the top view of the triangular magnetic alignments of Mn moments within each Mn-Sn Kagome plane. There are two types of the magnetic alignments observed in $Mn_3Sn$, denoted as AFM1 and AFM2. **d** The calculated magnetic spin Hall conductivity $\sigma_{zx}^z$ and $\sigma_{zy}^z$ in $Mn_3Sn$. Left panel shows the $\sigma_{zx}^z$ and $\sigma_{zy}^z$ as a function of energy for AFM1 and AFM2. Right panel shows the sign change of $\sigma_{zx}^z$ and $\sigma_{zy}^z$ at $E_F$ when the magnetic moments in AFM1 and AFM2 are reversed by in-plane magnetic fields. The finite $\sigma_{zx}^z$ and $\sigma_{zy}^z$ indicate $Mn_3Sn$ can be a spin source for the device shown in **b** to support an out-of-plane anti-damping torque.

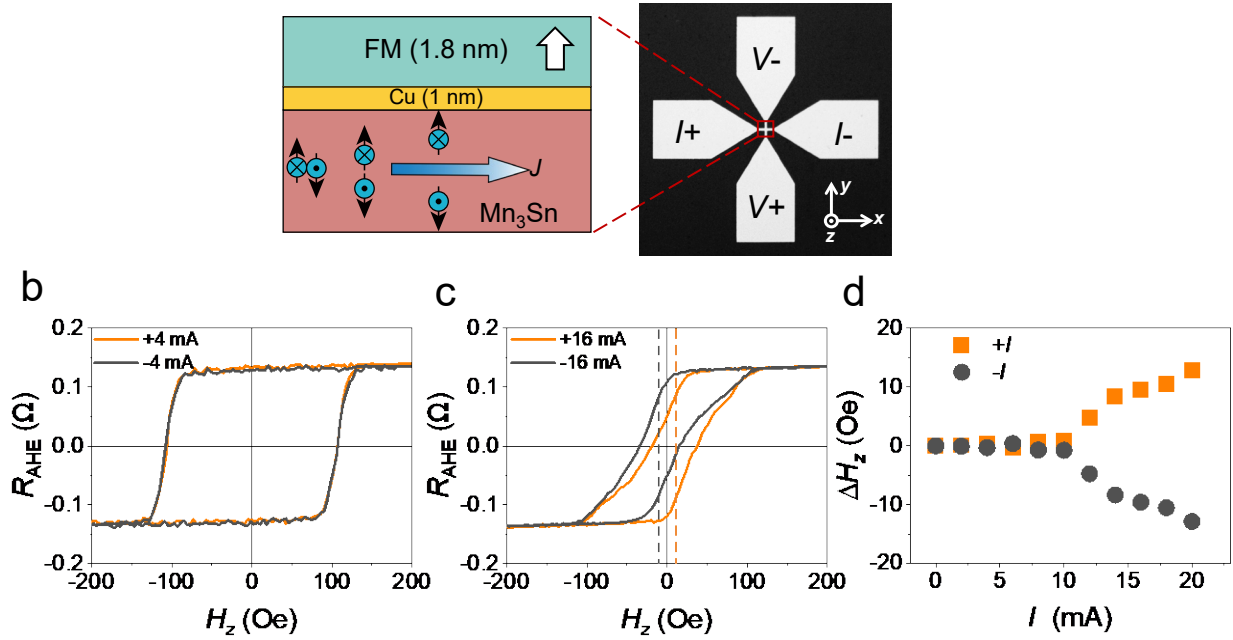

**Figure 2: *Z*-polarized spin current generated by Mn₃Sn thin film. a** The schematic of Mn$_3$Sn (7)/Cu (1)/FM (1.8) stack (left) and optical image of the device using for electrical transport measurements (right). The spins with both ±$y$ and ±$z$ polarizations generated by bottom Mn$_3$Sn thin film will act on the ferromagnetic layer and induce spin orbit torques simultaneously. **b, c** $R_{AHE}$ vs. $H_z$ curve when the bias currents are ±4 mA and ±16 mA. **d** A summary of the shift (Δ$H_z$) at different bias currents (*I*). The threshold *I* to cause a shift in AHE curve is about 10 mA. +*I* will shift the AHE curve to the +*x* while -*I* leads to the opposite shift.

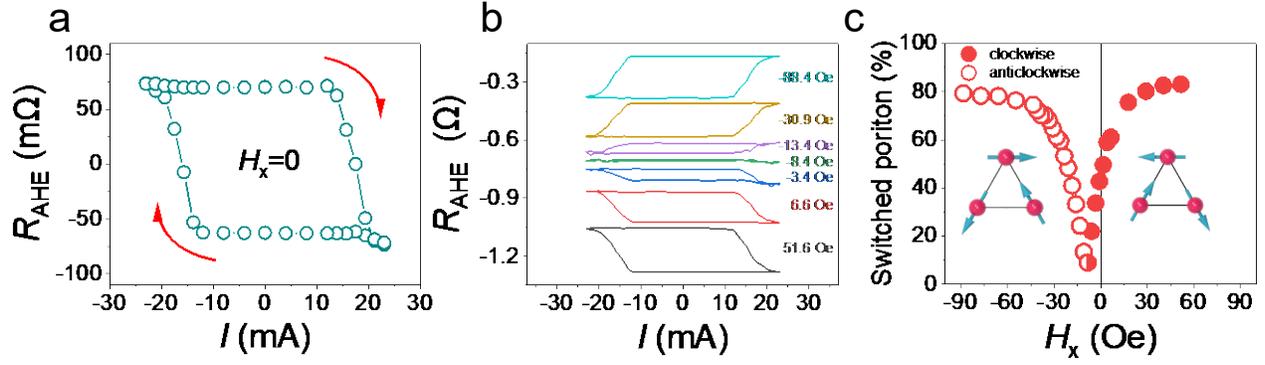

**Figure 3: External magnetic field tuning polarity of current induced magnetization switching. a** Current induced magnetization switching with clockwise polarity in the absence of an external magnetic field. **b** The switching curve under different external magnetic fields from negative to positive. **c** The evolution of the switching polarity and switching ratio under different magnetic fields. Here the direction of magnetic field is in-plane and parallel to the current. The two opposite $Mn_3Sn$ domains contribute opposite *z* spins and thus induce clockwise and anticlockwise switching polarity as indicated as solid and hollow dots.

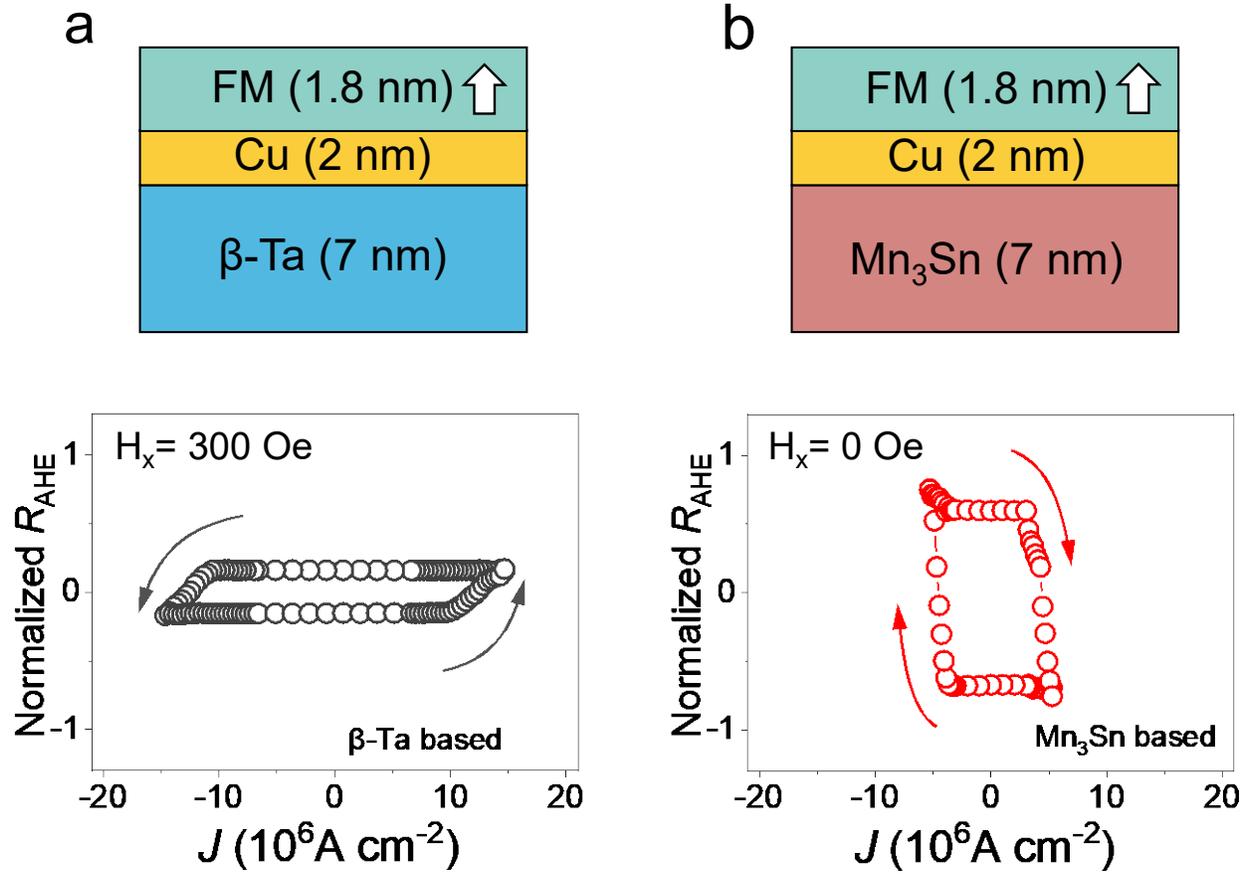

**Figure 4: High MSHE based SOT efficiency with the assistance of *z* spin polarization. a** The conventional SHE based SOT device with a structure of *β*-Ta (7)/Cu (2)/FM (1.8) and its maximum current induced magnetization switching curve at the external magnetic field of 300 Oe. **b**. The novel MSHE based SOT device with a structure of Mn$_3$Sn (7)/Cu (2)/FM (1.8) and its current induced magnetization switching curve of Mn$_3$Sn based device with the absent of external magnetic field. Note that the current density is calculated by considering the shunting effect of Cu and FM layer in both devices.